\documentclass[journal=nalefd,manuscript=letter,layout=twocolumn]{achemso}
\usepackage{amsmath}
\usepackage{blindtext}
\usepackage[version=3]{mhchem} 
\usepackage{upgreek}
\usepackage[utf8]{inputenc}
\usepackage[pdftex,unicode=true,bookmarks=false,breaklinks=false,pdfborder={0 0 1},colorlinks=true]{hyperref}
\hypersetup{linkcolor=blue,citecolor=blue,urlcolor=blue}

\newcommand{\celsius}{$~^{\circ}$C\xspace}

\author{Hanno Küpers}
\email{kuepers@pdi-berlin.de}
\affiliation{Paul-Drude-Institut für Festkörperelektronik, Hausvogteiplatz 5--7, 10117 Berlin, Germany}

\author{Ryan B. Lewis}
\affiliation{Paul-Drude-Institut für Festkörperelektronik, Hausvogteiplatz 5--7, 10117 Berlin, Germany}

\author{Abbes Tahraoui}
\affiliation{Paul-Drude-Institut für Festkörperelektronik, Hausvogteiplatz 5--7, 10117 Berlin, Germany}

\author{Matthias Matalla}
\affiliation{Ferdinand-Braun-Institut, Leibniz-Institut für Höchstfrequenztechnik, Gustav-Kirchhoff-Strasse 4, 12489 Berlin, Germany}

\author{Olaf Krüger}
\affiliation{Ferdinand-Braun-Institut, Leibniz-Institut für Höchstfrequenztechnik, Gustav-Kirchhoff-Strasse 4, 12489 Berlin, Germany}

\author{Faebian Bastiman}
\affiliation{Paul-Drude-Institut für Festkörperelektronik, Hausvogteiplatz 5--7, 10117 Berlin, Germany}

\author{Henning Riechert}
\affiliation{Paul-Drude-Institut für Festkörperelektronik, Hausvogteiplatz 5--7, 10117 Berlin, Germany}

\author{Lutz Geelhaar}
\affiliation{Paul-Drude-Institut für Festkörperelektronik, Hausvogteiplatz 5--7, 10117 Berlin, Germany}

\title{Diameter evolution of selective area grown Ga-assisted GaAs nanowires}


\begin{document}

\begin{abstract}
We present a novel two-step approach for the selective area growth (SAG) of GaAs nanowires (NWs) by molecular beam epitaxy which has enabled a detailed exploration of the NW diameter evolution. In the first step, the growth parameters are optimized for the nucleation of vertically-oriented NWs. In the second step, the growth parameters are chosen to optimize the NW shape, allowing NWs with a thin diameter (45~nm) and an untapered morphology to be realized. This result is in contrast to the commonly observed thick, inversely tapered shape of SAG NWs. We quantify the flux dependence of radial vapour-solid (VS) growth and build a model that takes into account diffusion on the NW sidewalls to explain the observed VS growth rates. Combining this model for the radial VS growth with an existing model for the droplet dynamics at the NW top, we achieve full understanding of the diameter of NWs over their entire length and the evolution of the diameter and tapering during growth. We conclude that only the combination of droplet dynamics and VS growth results in an untapered morphology. This result enables NW shape engineering and has important implications for doping of NWs.
\end{abstract}

In recent years, numerous electronic and optoelectronic devices based on GaAs nanowires (NWs) have been demonstrated, including light-emitting diodes \cite{Tomioka_nl_2010}, lasers \cite{Mayer2013}, and photovoltaic cells \cite{LaPierre2013a}. These structures can be grown directly on Si substrates, enabling the integration of direct bandgap III-V devices with Si technology. For many device structures it is crucial to control the diameter and shape of the NWs. In the Ga-assisted vapour-liquid-solid~(VLS) growth of GaAs NWs, a Ga droplet acts as a collector for Ga and As and no external catalyst material is necessary\cite{Colombo_prb_2008,Jabeen_nt_2008}. However, the dynamics of the Ga droplet during NW growth can easily lead to a diameter variation of the forming NW, resulting in either positively or negatively (inversely) tapered NWs. Recently, theoretical models have been established to describe the shape of such NWs based on the droplet dynamics\cite{Tersoff2015a,Dubrovskii2016a}. However, in addition to VLS growth at the droplet, direct vapour-solid (VS) growth on the side facets is regularly observed, which also influences the NW diameter,\cite{Colombo_prb_2008,Sartel_jcg_2010,Rieger2012,LaPierre2013,Munshi2014}  but this phenomenon lacks a comprehensive description.

For many applications the positioning of individual NWs on the substrate is essential to gain control over the performance of single NWs and collective photonic effects\cite{Heiss2014}. Consequently, the selective area growth (SAG) of Ga-assisted NWs in molecular beam epitaxy (MBE) has been the focus of intense investigation in recent years \cite{Bauer2010,Plissard2010}. In the common approach, arrays of holes are defined by lithography into thermal SiO$_2$ layers. Due to the low sticking on the oxide surface, the growth is restricted to the oxide-free holes. The resulting NWs typically exhibit a larger diameter and more tapering compared to NWs grown on unpatterned Si substrates covered with native oxide\cite{Bastiman2016}, which is undesirable for most applications. The large diameter and negatively tapered morphology are assumed to result from the different growth conditions needed for NW growth on patterned substrates leading to an enlargement of the Ga droplet \cite{Tersoff2015a}.

In this work, we develop a two-step growth procedure to decouple NW nucleation in the mask holes from NW elongation. NWs are nucleated at a low V/III ratio to maximizie the vertical yield (the fraction of holes occupied by vertical NWs). After nucleation, the growth conditions are adjusted to tailor the NW morphology. Using this approach, we realize high vertical yields of thin and untapered NWs with lengths of several $\upmu \text{m}$. With this versatile growth procedure, we explore the elongation phase in detail. We find that radial VS growth has a strong impact on the final diameter and shape of the NW. We present a model that explains the observed radial growth and is consistent with the understanding of diffusion processes that are responsible for the axial growth. Finally, we combinine our VS growth model with an existing model for VLS growth by Tersoff \cite{Tersoff2015a}. Thereby we can successfully describe the shape of complete NWs and its evolution during elongation.


\begin{figure*}
		\centering
		\includegraphics[width=\textwidth]{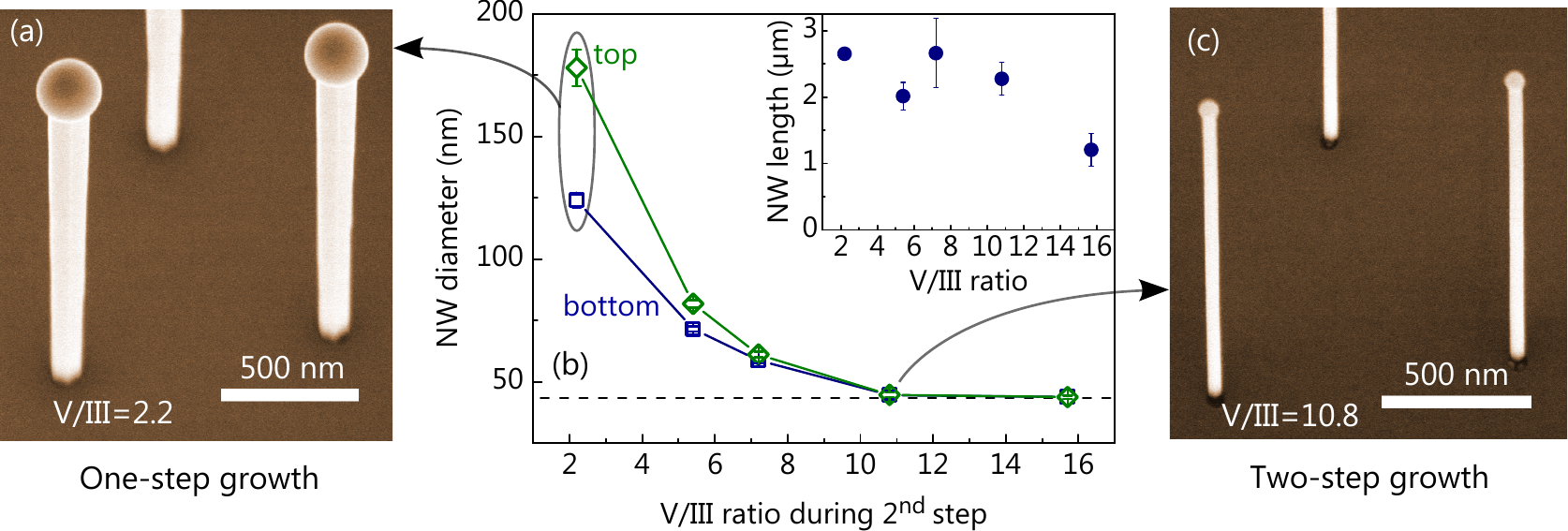}
		\caption{(a) SEM image of NWs grown with a constant V/III flux ratio of 2.2 for 30~min. The sample is tilted by 25$^\circ$ from the substrate normal.
		(b) NW diameter at top (green diamonds) and bottom (blue squares) for different V/III flux ratios during the second growth step. Inset: NW length for varying V/III ratio during the second step. The NW length is independent of the V/III ratio (Ga flux) except for the highest V/III ratio, where the NWs are significantly shorter due to the termination of the VLS growth.
		(c) SEM image of NWs grown with a two-step growth approach under a second-step V/III flux ratio of 10.8, exemplifying the efficacy of the growth approach. The sample is tilted by 25$^\circ$.
				}
		\label{fig:Morphology2step}
	\end{figure*}	

For SAG of Ga-assisted GaAs NWs it was shown previously that low V/III ratios are necessary for a high vertical yield\cite{Plissard2011,Munshi2014}. Such conditions lead to locally Ga-rich conditions enabling the VLS growth of NWs. Our own optimization of the vertical yield confirms this trend (see supporting information), and a sample that was grown under a V/III ratio of 2.2, which is in the optimum region, is shown in Figure~\ref{fig:Morphology2step}~(a). The NW diameter is 160~nm at the top and tapering is pronounced at -2.5\% (linearized tapering $\frac{d_{\text{bot}}-d_{\text{top}}}{l}$, where $d_{\text{bot}}$ and $d_{\text{top}}$ are the NW diameter at bottom and top, respectively, and $l$ is the NW length\cite{Colombo_prb_2008}). We also note that higher resolution images show steps on the NW side facets. Diameters reported in literature for SAG are typically larger than the diameter for NWs grown on unpatterned substrates. For SAG the reported NW diameters are typically well above 60~nm for NW lengths of several $\upmu$m\cite{Gibson2013b,Munshi2014}, whereas for growth on native oxide diameters down to 30~nm have been reported \cite{Colombo_prb_2008,Bastiman2016}. Furthermore, most reported SAG NWs were negatively tapered\cite{Bauer2010,LaPierre2013,Rudolph2014,Munshi2014}. Both the large diameter and the tapering are a consequence of the lower V/III ratio, resulting in a large VLS droplet\cite{Colombo_prb_2008,Tersoff2015a}. 

The different requirements for vertical growth and thin diameters imply that both features cannot be achieved with a single set of growth conditions. To overcome this difficulty, we developed a growth approach with two sets of growth parameters for the different phases of growth: The first step provides growth conditions necessary to achieve a high vertical yield. These are a 90~s pre-deposition phase for Ga droplet formation, and subsequently GaAs growth with a low V/III ratio of 2.2. This step lasts 300~s, which is long enough to establish the stable VLS growth of NWs which reach a length of about 300~nm and have a diameter of 30~nm. In the second step the V/III ratio is increased by decreasing the Ga flux by closing one of the two Ga cells. This approach allows the V/III ratio to be increased without increasing the As flux, which would change the elongation rate (elongation varies linearly with As flux \cite{Colombo_prb_2008}. The total growth time for these two-step samples was 30~min. Figure~\ref{fig:Morphology2step}~(b) presents the diameter at the top (green diamonds) and bottom (blue squares) of the NWs for varying V/III ratios during the second step. Both diameters decrease monotonically with increasing V/III ratio, becoming constant at a value of 45~nm for V/III ratios above 10.8. The inset of Figure~\ref{fig:Morphology2step}~(b) shows the NW length for varying V/III ratio in the second step. The length is roughly constant as expected for a constant growth time and As flux\cite{Colombo_prb_2008,Rieger2012,Bastiman2016}. However, the NWs grown with highest V/III flux ratio of 15.7 are significantly shorter. These NWs exhibit a small Ga droplet with a low contact angle, indicating that the VLS growth is ceasing due to the high V/III ratio. For the highest V/III ratios the NWs have an untapered morphology.

Figure~\ref{fig:Morphology2step} (c) shows a micrograph of the sample grown with a V/III ratio of 10.8 during the second step. This micrograph illustrates a thin and untapered morphology that has not been shown previously for selective area grown GaAs NWs of this length. Furthermore, the sample exhibits a vertical yield of 55\%, which is comparable to our optimized one-step yield. This result illustrates the efficacy of the two-step approach, which allows for the optimal V/III ratio both for nucleation and NW morphology.


\begin{figure*}
		\centering
		\includegraphics[width=\textwidth]{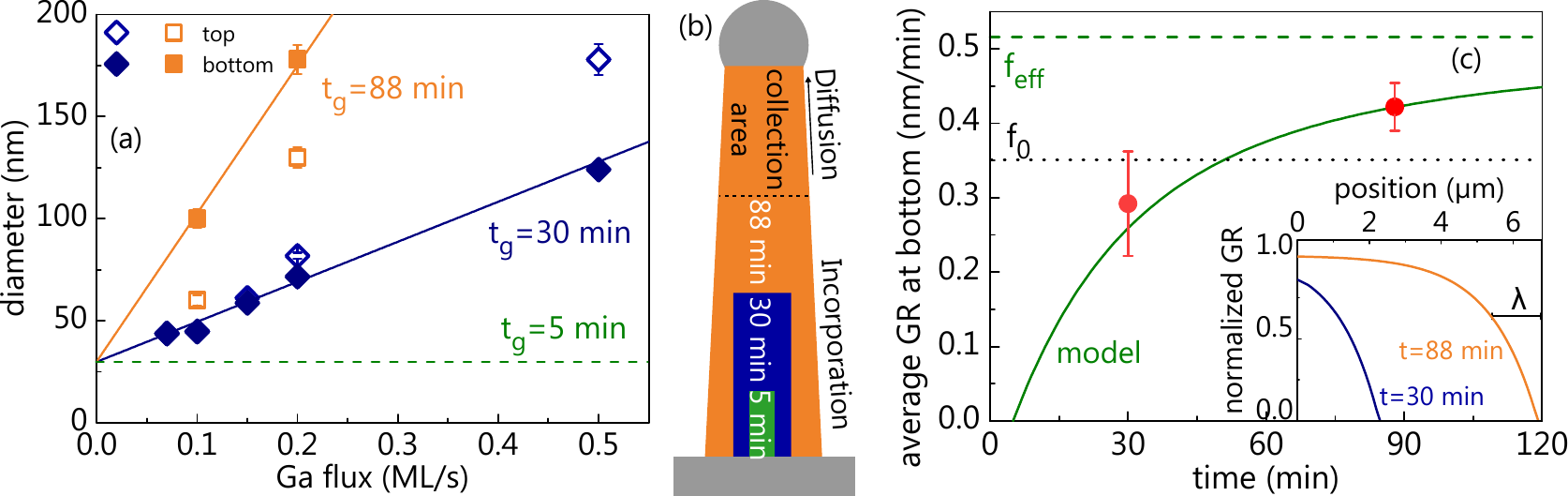}
		\caption{(a) Diameter at the bottom (full symbols) and top (open symbols) of the NWs for varying Ga flux. The blue diamonds represent the same data as in Figure~\ref{fig:Morphology2step}~(b) (30~min growth time) and the orange squares represent samples grown for 88~min. The lines are guides to the eye for the bottom diameters of the two series. The green dashed line represents the bottom diameter at the beginning of the second step.
		(b) Sketch of the NWs grown for different times, visualizing the VLS and VS growth processes:~Within the collection area, most adatoms diffuse to the droplet, whereas below the collection area most adatoms are incorporated into the NW sidewall.
		(c) Average radial growth rate at the bottom of the NW calculated by equation~\ref{eq:GRav_exp} for different times. The green line is a fit of equation~\ref{eq:effGR} to the data. The black dotted line represents the nominal flux $f_0$ from the effusion cell and the green dashed line the effective flux $f_\text{eff}$ as determined by our model. Inset: Normalized differential radial growth rate as a function of position along the NW after 30~min of growth (blue line) and 88~min (orange line), as calculated by equation~\ref{eq:growth rate}.
		}
		\label{fig:RadialGrowthRate}
	\end{figure*}	

In this study, all two-step samples start from a similar NW base which is grown within the first step. Thus, the diameter at the bottom of the NWs at the beginning of the second step is the same for all samples (30~nm) and it is unaffected by the droplet dynamics during the second step. Therefore, the variation of the bottom diameter in Figure~\ref{fig:Morphology2step}~(b) can only be caused by direct VS growth on the side-facets. In order to understand this phenomenon in more detail we plot the bottom diameters from Figure~\ref{fig:Morphology2step}~(b) as a function of Ga flux in Figure~\ref{fig:RadialGrowthRate}~(a), making it more convenient to analyze VS growth. Additionally, we show data for samples with a longer growth time (88~min). For both series the bottom diameter increases linearly with increasing Ga flux (lines are guides to the eye), a clear indication of VS growth. 

Figure~\ref{fig:RadialGrowthRate}~(a) also shows the top diameter for the short and long NW series (open symbols), allowing the NW shape to be compared for the two growth times. Although NWs grown with Ga flux of 0.1~ML/s show no tapering after 30~min, the top diameter after 88~min is much smaller than the bottom diameter. In other words, the tapering of the NW changes throughout growth and the untapered NW shape holds only for a certain growth time. The evolution of the NW shape with time is visualized in Figure~\ref{fig:RadialGrowthRate}~(b). 

To understand the above results, we first want to concentrate on the VS radial growth that happens at the bottom of the NWs. Focusing on the sample series grown with a V/III ratio of 10.8, which leads to the smallest diameters, we calculate time-averaged radial growth rates $GR_{\text{av}}$ as
\begin{equation}
\label{eq:GRav_exp}
GR_{\text{av}}=(r_{\text{g}}-r_1)/(t_\text{g}-t_1)~,
\end{equation}
with the NW radius at the bottom $r_{\text{g}}$ after growth time $t_\text{g}$, and the NW radius at the bottom after the first step $r_1=30$~nm at $t_1=5$~min. These average growth rates are useful for considering only the growth during the second step. Figure~\ref{fig:RadialGrowthRate}~(c) shows the calculated growth rates for the respective growth times based on the experimental diameters in Figure~\ref{fig:RadialGrowthRate}~(a). The average growth rate increases with growth time, and for 88~min it is larger than the growth rate corresponding to the directly impinging Ga flux from the effusion cell taking into account substrate rotation and angle between substrate normal and effusion cell. This finding indicates that a secondary flux exists, most probably the re-evaporation of atoms from the oxide surface to the NW side-facets \cite{Dalacu2009,Kelrich2013,Gibson2014}.


A central factor of VLS growth of NWs is a large diffusion length on the NW sidefacets \cite{Jabeen_nt_2008,Dubrovskii2015b}. Ga atoms that impinge on the NW sidefacet diffuse along the NW axis to reach the Ga droplet where they incorporate into the lattice at the liquid-solid interface. However, this supply by diffusion takes place mostly within a diffusion length from the droplet as indicated by the `collection area' in Figure~\ref{fig:RadialGrowthRate}~(b). For long NWs the bottom might be too far away for Ga atoms to reach the droplet. If we neglect desorption of Ga atoms on the NW sidewalls, which is a reasonable assumption at the growth temperature of 630\celsius \cite{Ralston}, all Ga atoms that do not contribute to the diffusion supply must incorporate into the NW sidewall. For the quantitative description of the diffusion we solve the one-dimensional diffusion equation\cite{Fick1851}. The adatom density is given by

\begin{equation}
\label{eq:density}
n(z,t)=\tau f_{\text{eff}} \left(1 - exp\left(- \frac{l(t)-z}{\lambda}\right) \right)~,
\end{equation}

with the surface lifetime before incorporation $\tau$ (neglecting desorption), the total impinging Ga flux $f_{\text{eff}}$, the time-dependent NW length $l(t)$, the position on the NW axis $z$, and the diffusion length $\lambda$\cite{Tersoff2015a}. We calculate the radial growth rate at a specific point $z$ as

\begin{equation}
\label{eq:growth rate}
GR(z,t) = \begin{cases}
n(z,t)/\tau &\text{for $z\leq l(t)$}\\
0 &\text{for $z > l(t)$~.}
\end{cases}
\end{equation}

This equation, using a dimension-less $f_{\text{eff}}=1$ and $\lambda =1.2~\upmu \text{m}$, is plotted in the inset of Figure~\ref{fig:RadialGrowthRate}~(c) for two different times, corresponding to different NW lengths. The local radial VS growth rate strongly decreases near the droplet within a length comparable to the diffusion length, representing the collection area. Consequently, the radial VS growth rate at the bottom of the NW changes with time.

From equation \ref{eq:growth rate} we calculate the radially grown thickness at position $z$ for the duration $t_1$ to $t_\text{g}$ by integrating $GR(z,t)$ as
\begin{equation}
\label{eq:grownthickness}
r_{\text{VS}}(z,t_\text{g})=\int_{t_{1}}^{t_{\text{g}}} GR(z,t) dt~. 
\end{equation}
and therefore the average growth rate follows as
\begin{equation}
\label{eq:effGR}
GR_{\text{av}}(z,t_\text{g})=\frac{1}{t_\text{g}-t_1}\int_{t_{1}}^{t_{\text{g}}} GR(z,t) dt~. 
\end{equation}
The model depends only on two free parameters, $f_{\text{eff}}$ and $\lambda$. The time dependent length is calculated from the mean NW lengths in Figure~\ref{fig:Morphology2step}~(b), by assuming a constant growth rate, yielding $GR_{\text{ax}}~=~76~\text{nm/min}$.

A fit of equation \ref{eq:effGR} to the experimental data is shown by the green line in Figure~\ref{fig:RadialGrowthRate}~(c). Here, we use $t_1=5~\text{min}$. An impinging flux $f_{\text{eff}}$ of 0.52~nm/min (shown as green dashed line) is obtained, which amounts to 150\% of the nominal impinging flux $f_0$ (black dotted line). This value is consistent with results reported for re-evaporation of As from the substrate and NW sidewalls \cite{Ramdani2013} and confirms the hypothesis that Ga is re-evaporated from the substrate surface. Also the lack of growth on the oxide surface confirms that Ga desorbs from the oxide. A diffusion length of 1.2~$\upmu$m is obtained from the fit, which is a reasonable value for this system and comparable to earlier results\cite{Dubrovskii2015b}.

We checked the consistency of our model with the experimentally observed NW axial growth. From SEM images, we estimate the volume of GaAs grown in the VLS mode between 30~min and 88~min to be $V_{\text{grown}}=8.64\cdot10^6~\text{nm}^3$. The amount of Ga collected by the droplet on the NW sidewall based on our model is $V_{\text{col}}=8.05\cdot10^6~$nm$^3$ as GaAs equivalent volume. The result that both volumes are comparable confirms the consistency of our model for the radial growth with the measured axial growth rate (Details in supporting information).

\begin{figure*}
		\centering
		\includegraphics[width=0.85\textwidth]{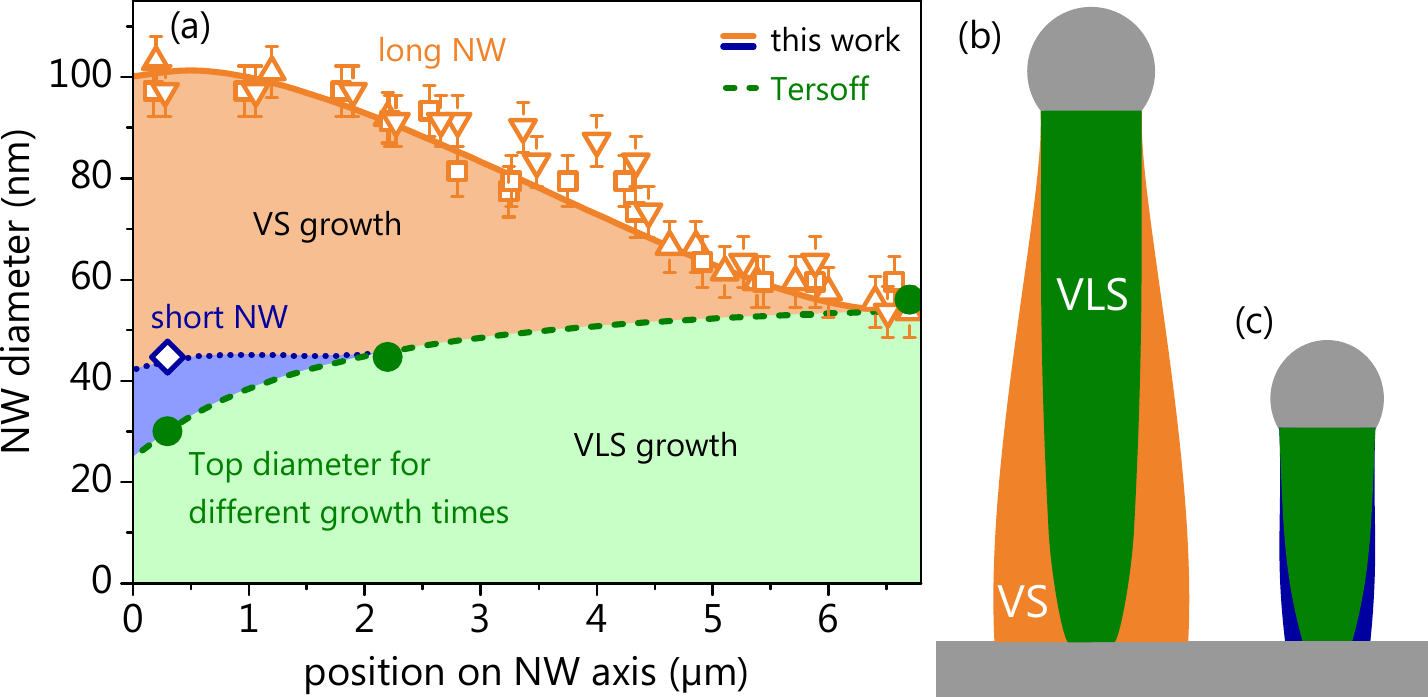}
		\caption{(a) NW diameter as a function of position along the NW axis for NWs grown for 5, 30 and 88~min. The full green circles show the top diameter for three different times corresponding to different lengths. The dashed green line is a fit by equation~\ref{eq:diffEq} describing droplet dynamics~\cite{Tersoff2015a}. The open orange triangles and squares represent the diameter at different points of three long NWs. The diameter as calculated by equation~\ref{eq:complete} is shown as the orange line. The dotted blue line shows the calculated values for the short NWs and is in agreement with the experimental value at the NW bottom shown as open blue diamond. The calculated shapes of the long and short NWs are shown schematically in (b) and (c), respectively.
		}
		\label{fig:RadialGrowth}
	\end{figure*}

So far, we have established a model for the radial VS growth and have used it to explain the diameter widening observed at the bottom of the NWs. For a full description of the shape of a NW, both droplet dynamics (VLS) and radial growth (VS) need to be accounted for over the entire NW length. Tersoff has built a model for the self-assisted VLS NW growth based on rate equations in order to explain the tapering due to the dynamics of the droplet \cite{Tersoff2015a}. According to that model, VLS growth can be stable for a large range of V/III ratios. The V/III ratio in combination with a natural length scale --- the diffusion length $\lambda$ --- and a geometrical factor --- the ratio of droplet height and NW radius $\eta=h/r$ --- determine the dynamics of the NW diameter. However, in this model radial VS growth is neglected.

The model by Tersoff describing the radiusduring VLS growth $r_{\text{VLS}}$ over NW length $z$ is described by the equation:
\begin{equation} \label{eq:diffEq}
\begin{split}
\frac{dr_{\text{VLS}}}{dz} =& \\
 \frac{\frac{2 \Omega_\text{L}}{\Omega_\text{x}}}{\eta(3+\eta^2)} & \left[  \frac{F_3}{F_5} \left(  1+\frac{\lambda}{(1+\eta^2)r_{\text{VLS}}} \right)  -1 \right]~,
\end{split}
\end{equation}
with $\Omega_\text{L}$ the atomic volume of liquid Ga, $\Omega_\text{x}$ the volume per two-atom unit of the crystal, and $F_3$ and $F_5$ the group III and V fluxes.

As a starting point for the full description of the NW shape, we consider the evolution of the top diameter with growth time. The top diameter is not affected by radial VS growth, and thus, these values are described well by Tersoff's model. The green circles in Figure~\ref{fig:RadialGrowth} (a) show the top diameter of the three samples grown for different times with the same growth conditions, plotted as a function of length. The dashed green line represents a fit of the solution of equation \ref{eq:diffEq} to the experimental data. The free fit parameters are a droplet shape factor $\eta$ of 3.35 and an effective V/III ratio $F_5/F_3$ of 5 ($\Omega_\text{L} / \Omega_\text{x} \approx 0.42$ for GaAs). The diffusion length was set to 1.2~$\upmu \text{m}$ as determined above.

The orange triangles and squares in Figure~\ref{fig:RadialGrowth}~(a) represent the diameter along the axis of three 7~$\upmu$m long NWs as measured by SEM after 88~min of growth. In striking contrast to the evolution of the top diameter, these long NWs exhibit a positively tapered shape, resulting from the radial VS growth. In order to describe the former dependence, we amended Tersoff's model to include direct radial VS growth. We simply add the contributions for VLS growth (solution of equation~\ref{eq:diffEq}) and VS growth (equation~\ref{eq:grownthickness}) because radial growth due to droplet dynamics and VS growth are independent of each other:
\begin{equation}
\label{eq:complete}
r_{\text{tot}} = r_{\text{VLS}} + r_{\text{VS}}~.
\end{equation}
The result of this equation --- using the fit parameters from the evaluation of the top diameters for the VLS contribution (fit to equation~\ref{eq:diffEq}) and from the fit of the VS model to the average radial VS growth rates at the bottom (fit to equation~\ref{eq:effGR}) --- is shown as the orange line in Figure~\ref{fig:RadialGrowth} (a). This model describes the experimental data well. At the bottom a rather flat part is present in both the model and the experimental data. Here, the droplet dynamics lead to strong negative tapering, as seen in the green curve. This effect is compensated by the VS growth, yielding a flat bottom part. The upper part of the NW exhibits a positive tapering as it is dominated by VS growth. The amount of NW material resulting from axial VLS and radial VS growth is indicated by the green and orange areas, respectively. A visualization of the calculated NW shape with the parts due to VLS and VS growth is shown in Figure~\ref{fig:RadialGrowth}~(b). We emphasize the strong agreement between the experimental data and our model. The model curves are not fits to the data, but are calculations using the fit parameters gained from the analysis of the VS growth at the NW bottom by equation~\ref{eq:effGR}. The good agreement between experimental data and model seen in Figure~\ref{fig:RadialGrowth}~(a) demonstrates the validity of our model. 

We use our comprehensive model to describe the shape of the short NWs, which is shown as the blue line in Figure~\ref{fig:RadialGrowth}~(a). In agreement with our experimental results, the shape is essentially untapered. Our model reveals that the untapered morphology as shown in Figure~\ref{fig:RadialGrowth}~(c) is a consequence of the droplet dynamics and the VS growth compensating each other. Thus, by balancing droplet dynamics and VS growth, straight NWs can be achieved but only for a certain range of NW lengths.


We have shown that a substantial part of the NW volume is grown by direct radial VS growth. For our long NWs the volume fraction resulting from VS growth amounts to 63\%. This finding has several implications.
It was suggested previously that during the Ga-assisted VLS growth of GaAs NWs the incorporation of dopants is mostly due to radial growth and in-diffusion on the sidefacets while the droplet is only a small channel for the incorporation of dopant atoms\cite{Casadei2012}. If we exclude diffusion and desorption of dopant atoms, the dopant density is determined by the GaAs growth rate on the sidewall. As we showed, this growth rate is smallest just below the droplet leading to a high density of dopant atoms. Correspondingly, the VLS grown core will be surrounded by a VS grown shell with a gradient in doping density.

Finally, this study used SAG in order to have a better control over NW growth. Our study is the first experimental confirmation of the shape of GaAs NWs described by a theoretical model that describes the atomic processes on the NW sidefacet. At the same time our comprehensive model is based on a consideration of the incorporation pathways and is therefore not restricted to the SAG of NWs. As droplet dynamics and radial VS growth are fundamental processes for VLS NW growth, the basic ideas of our model can also be applied to other material systems.


In conclusion, our novel two-step growth approach enables simultaneous optimization of vertical yield and NW morphology. We show that increasing the V/III ratio during the second step leads to a clear reduction of the NW diameter, and the realization of untapered NWs with diameters of 45~nm at a length of 2.5~$\upmu$m. This achievement is the ideal basis for the growth of strained shells on such thin cores. Despite the high growth temperature of 630\celsius , radial VS growth is significant and leads to an increase in diameter during growth.

On the basis of these findings we developed a comprehensive growth model that takes into account both factors influencing the NW diameter:~droplet dynamics and radial VS growth. This model successfully describes the evolution of the NW morphology throughout the growth. An untapered NW shape requires a balance between droplet dynamics and radial VS growth, which can be achieved for given growth conditions only for a certain growth time and hence length. Very importantly, even in untapered NWs a large part of the NW volume can result from radial VS growth. This insight is expected to have consequences for doping. Our understanding of the factors that determine the NW shape now enables its prediction and the engineering of different NW shapes.

\section{Experiments and methods}

Samples were grown by MBE on Si(111) P-doped (n-type) substrates covered by a patterned oxide mask. The MBE system contains two effusion cells for Ga and a valved cracker source for supply of As$_2$. For the measuring the substrate temperature an optical pyrometer was used, which was calibrated to the oxide desorption temperature of GaAs(100) \cite{Bastiman2016}. Before starting the growth, the substrates were annealed at about 680\celsius in the growth chamber for 10 minutes. Then, the temperature was changed to the growth temperature of 630\celsius . In order to form droplets in the mask holes, Ga was deposited for 90~s at a flux of 0.5~ML/s using both Ga cells. Subsequently, NW growth was initiated by the simultaneous supply of Ga and As$_2$. For one-step growth experiments the V/III flux ratio was varied by changing the As flux at a constant Ga flux of 0.5~ML/s. The growth time was 15--30~min. For the two-step growth the As flux was fixed at 1.1~ML/s (V/III=2.2 in first step). After 5~min of growth the second growth step was started by closing one of the two Ga cells in order to reduce the Ga flux (V/III=2.2--15.8 in second step). The growth was stopped after 25~min by closing all sources and ramping the substrate to 100\celsius at 2\celsius /s.

The oxide mask comprises fields with hexagonal patterns of holes with 1~$\upmu$m pitch and a hole diameter of 40--50~nm. The hole pattern was written by electron beam lithography (EBL) on 2" and 3" wafers covered with a 15--20~nm thick thermal SiO$_2$. The mask was etched into the oxide by reactive ion etching using CHF$_3$. Subsequently, the samples were cut into square pieces with an edge length of 10~mm and then cleaned by organic solvents, oxygen plasma and UV ozone. Just before loading samples into the MBE system the native oxide in the holes was etched for 60~s in a 1$\%$ solution of hydrofluoric acid (HF) and rinsed in ultrapure water (conductivity $\rho$=0.005$~\frac{\mathrm{\upmu S}}{\mathrm{cm}}$ and total organic carbon (TOC) concentration $\leq 1$~ppb). After the cold water rinse the substrates were boiled in ultrapure water for 10~min. Finally, the samples were blow dried with nitrogen and loaded into the MBE load lock. Further details about the substrate preparation can be found elsewhere\cite{SAGtech}.

The NW morphology was measured by scanning electron microscopy (SEM) from either side-view images or top-view images inclined by 25$~^{\circ}$ from the substrate normal. The error bars in the view-graphs reflect standard deviation for ensemble measurements and error of measurement for single NWs.

\begin{acknowledgement}

This work was supported by the Deutsche Forschungsgemeinschaft (DFG) under grant Ge2224/2, R.B.L. acknowledges funding from the Alexander von Humboldt Foundation. We are grateful to Anne-Kathrin Bluhm for acquiring SEM images and to Michael Höricke and Carsten Stemmler as well as Arno Wirsig for technical support at the MBE system. We appreciate the critical reading of the manuscript by Patrick Vogt.
\end{acknowledgement}

\bibliography{libSAG}

\providecommand{\latin}[1]{#1}
\makeatletter
\providecommand{\doi}
  {\begingroup\let\do\@makeother\dospecials
  \catcode`\{=1 \catcode`\}=2\doi@aux}
\providecommand{\doi@aux}[1]{\endgroup\texttt{#1}}
\makeatother
\providecommand*\mcitethebibliography{\thebibliography}
\csname @ifundefined\endcsname{endmcitethebibliography}
  {\let\endmcitethebibliography\endthebibliography}{}
\begin{mcitethebibliography}{28}
\providecommand*\natexlab[1]{#1}
\providecommand*\mciteSetBstSublistMode[1]{}
\providecommand*\mciteSetBstMaxWidthForm[2]{}
\providecommand*\mciteBstWouldAddEndPuncttrue
  {\def\EndOfBibitem{\unskip.}}
\providecommand*\mciteBstWouldAddEndPunctfalse
  {\let\EndOfBibitem\relax}
\providecommand*\mciteSetBstMidEndSepPunct[3]{}
\providecommand*\mciteSetBstSublistLabelBeginEnd[3]{}
\providecommand*\EndOfBibitem{}
\mciteSetBstSublistMode{f}
\mciteSetBstMaxWidthForm{subitem}{(\alph{mcitesubitemcount})}
\mciteSetBstSublistLabelBeginEnd
  {\mcitemaxwidthsubitemform\space}
  {\relax}
  {\relax}

\bibitem[Tomioka \latin{et~al.}(2010)Tomioka, Motohisa, Hara, Hiruma, and
  Fukui]{Tomioka_nl_2010}
Tomioka,~K.; Motohisa,~J.; Hara,~S.; Hiruma,~K.; Fukui,~T. \emph{Nano Letters}
  \textbf{2010}, \emph{10}, 1639--1644\relax
\mciteBstWouldAddEndPuncttrue
\mciteSetBstMidEndSepPunct{\mcitedefaultmidpunct}
{\mcitedefaultendpunct}{\mcitedefaultseppunct}\relax
\EndOfBibitem
\bibitem[Mayer \latin{et~al.}(2013)Mayer, Rudolph, Schnell, Morkötter,
  Winnerl, Treu, Müller, Bracher, Abstreiter, Koblmüller, and
  Finley]{Mayer2013}
Mayer,~B.; Rudolph,~D.; Schnell,~J.; Morkötter,~S.; Winnerl,~J.; Treu,~J.;
  Müller,~K.; Bracher,~G.; Abstreiter,~G.; Koblmüller,~G.; Finley,~J.~J.
  \emph{Nature communications} \textbf{2013}, \emph{4}, 2931\relax
\mciteBstWouldAddEndPuncttrue
\mciteSetBstMidEndSepPunct{\mcitedefaultmidpunct}
{\mcitedefaultendpunct}{\mcitedefaultseppunct}\relax
\EndOfBibitem
\bibitem[Lapierre \latin{et~al.}(2013)Lapierre, Chia, Gibson, Haapamaki,
  Boulanger, Yee, Kuyanov, Zhang, Tajik, Jewell, and Rahman]{LaPierre2013a}
Lapierre,~R.~R.; Chia,~A. C.~E.; Gibson,~S.~J.; Haapamaki,~C.~M.;
  Boulanger,~J.; Yee,~R.; Kuyanov,~P.; Zhang,~J.; Tajik,~N.; Jewell,~N.;
  Rahman,~K. M.~a. \emph{Physica Status Solidi - Rapid Research Letters}
  \textbf{2013}, \emph{7}, 815--830\relax
\mciteBstWouldAddEndPuncttrue
\mciteSetBstMidEndSepPunct{\mcitedefaultmidpunct}
{\mcitedefaultendpunct}{\mcitedefaultseppunct}\relax
\EndOfBibitem
\bibitem[Colombo \latin{et~al.}(2008)Colombo, Spirkoska, Frimmer, Abstreiter,
  and {Fontcuberta i Morral}]{Colombo_prb_2008}
Colombo,~C.; Spirkoska,~D.; Frimmer,~M.; Abstreiter,~G.; {Fontcuberta i
  Morral},~A. \emph{Physical Review B} \textbf{2008}, \emph{77}, 155326\relax
\mciteBstWouldAddEndPuncttrue
\mciteSetBstMidEndSepPunct{\mcitedefaultmidpunct}
{\mcitedefaultendpunct}{\mcitedefaultseppunct}\relax
\EndOfBibitem
\bibitem[Jabeen \latin{et~al.}(2008)Jabeen, Grillo, Rubini, and
  Martelli]{Jabeen_nt_2008}
Jabeen,~F.; Grillo,~V.; Rubini,~S.; Martelli,~F. \emph{Nanotechnology}
  \textbf{2008}, \emph{19}, 275711\relax
\mciteBstWouldAddEndPuncttrue
\mciteSetBstMidEndSepPunct{\mcitedefaultmidpunct}
{\mcitedefaultendpunct}{\mcitedefaultseppunct}\relax
\EndOfBibitem
\bibitem[Tersoff(2015)]{Tersoff2015a}
Tersoff,~J. \emph{Nano Letters} \textbf{2015}, \emph{15}, 6609--6613\relax
\mciteBstWouldAddEndPuncttrue
\mciteSetBstMidEndSepPunct{\mcitedefaultmidpunct}
{\mcitedefaultendpunct}{\mcitedefaultseppunct}\relax
\EndOfBibitem
\bibitem[Dubrovskii(2016)]{Dubrovskii2016a}
Dubrovskii,~V.~G. \emph{Journal of Crystal Growth} \textbf{2016}, \emph{440},
  62--68\relax
\mciteBstWouldAddEndPuncttrue
\mciteSetBstMidEndSepPunct{\mcitedefaultmidpunct}
{\mcitedefaultendpunct}{\mcitedefaultseppunct}\relax
\EndOfBibitem
\bibitem[Sartel \latin{et~al.}(2010)Sartel, Dheeraj, Jabeen, and
  Harmand]{Sartel_jcg_2010}
Sartel,~C.; Dheeraj,~D.~L.; Jabeen,~F.; Harmand,~J.-C. \emph{Journal of Crystal
  Growth} \textbf{2010}, \emph{312}, 2073--2077\relax
\mciteBstWouldAddEndPuncttrue
\mciteSetBstMidEndSepPunct{\mcitedefaultmidpunct}
{\mcitedefaultendpunct}{\mcitedefaultseppunct}\relax
\EndOfBibitem
\bibitem[Rieger \latin{et~al.}(2012)Rieger, Heiderich, Lenk, Lepsa, and
  Gr{\"{u}}tzmacher]{Rieger2012}
Rieger,~T.; Heiderich,~S.; Lenk,~S.; Lepsa,~M.~I.; Gr{\"{u}}tzmacher,~D.
  \emph{Journal of Crystal Growth} \textbf{2012}, \emph{353}, 39--46\relax
\mciteBstWouldAddEndPuncttrue
\mciteSetBstMidEndSepPunct{\mcitedefaultmidpunct}
{\mcitedefaultendpunct}{\mcitedefaultseppunct}\relax
\EndOfBibitem
\bibitem[Gibson and LaPierre(2013)Gibson, and LaPierre]{LaPierre2013}
Gibson,~S.; LaPierre,~R. \emph{physica status solidi (RRL) - Rapid Research
  Letters} \textbf{2013}, \emph{7}, 845--849\relax
\mciteBstWouldAddEndPuncttrue
\mciteSetBstMidEndSepPunct{\mcitedefaultmidpunct}
{\mcitedefaultendpunct}{\mcitedefaultseppunct}\relax
\EndOfBibitem
\bibitem[Munshi \latin{et~al.}(2014)Munshi, Dheeraj, Fauske, Kim, Huh,
  Reinertsen, Ahtapodov, Lee, Heidari, van Helvoort, Fimland, and
  Weman]{Munshi2014}
Munshi,~A.~M.; Dheeraj,~D.~L.; Fauske,~V.~T.; Kim,~D.~C.; Huh,~J.;
  Reinertsen,~J.~F.; Ahtapodov,~L.; Lee,~K.~D.; Heidari,~B.; van Helvoort,~A.
  T.~J.; Fimland,~B.~O.; Weman,~H. \emph{Nano letters} \textbf{2014},
  \emph{14}, 960--6\relax
\mciteBstWouldAddEndPuncttrue
\mciteSetBstMidEndSepPunct{\mcitedefaultmidpunct}
{\mcitedefaultendpunct}{\mcitedefaultseppunct}\relax
\EndOfBibitem
\bibitem[Heiss \latin{et~al.}(2014)Heiss, Russo-Averchi,
  Dalmau-Mallorqu{\'{i}}, T{\"{u}}t{\"{u}}nc{\"{u}}oğlu, Matteini,
  R{\"{u}}ffer, Conesa-Boj, Demichel, Alarcon-Llad{\'{o}}, and {Fontcuberta i
  Morral}]{Heiss2014}
Heiss,~M.; Russo-Averchi,~E.; Dalmau-Mallorqu{\'{i}},~a.;
  T{\"{u}}t{\"{u}}nc{\"{u}}oğlu,~G.; Matteini,~F.; R{\"{u}}ffer,~D.;
  Conesa-Boj,~S.; Demichel,~O.; Alarcon-Llad{\'{o}},~E.; {Fontcuberta i
  Morral},~A. \emph{Nanotechnology} \textbf{2014}, \emph{25}, 014015\relax
\mciteBstWouldAddEndPuncttrue
\mciteSetBstMidEndSepPunct{\mcitedefaultmidpunct}
{\mcitedefaultendpunct}{\mcitedefaultseppunct}\relax
\EndOfBibitem
\bibitem[Bauer \latin{et~al.}(2010)Bauer, Rudolph, Soda, {Fontcuberta i
  Morral}, Zweck, Schuh, and Reiger]{Bauer2010}
Bauer,~B.; Rudolph,~A.; Soda,~M.; {Fontcuberta i Morral},~A.; Zweck,~J.;
  Schuh,~D.; Reiger,~E. \emph{Nanotechnology} \textbf{2010}, \emph{21},
  435601\relax
\mciteBstWouldAddEndPuncttrue
\mciteSetBstMidEndSepPunct{\mcitedefaultmidpunct}
{\mcitedefaultendpunct}{\mcitedefaultseppunct}\relax
\EndOfBibitem
\bibitem[Plissard \latin{et~al.}(2010)Plissard, Dick, Larrieu, Godey, Addad,
  Wallart, and Caroff]{Plissard2010}
Plissard,~S.; Dick,~K.~a.; Larrieu,~G.; Godey,~S.; Addad,~A.; Wallart,~X.;
  Caroff,~P. \emph{Nanotechnology} \textbf{2010}, \emph{21}, 385602\relax
\mciteBstWouldAddEndPuncttrue
\mciteSetBstMidEndSepPunct{\mcitedefaultmidpunct}
{\mcitedefaultendpunct}{\mcitedefaultseppunct}\relax
\EndOfBibitem
\bibitem[Bastiman \latin{et~al.}(2016)Bastiman, K{\"{u}}pers, Somaschini, and
  Geelhaar]{Bastiman2016}
Bastiman,~F.; K{\"{u}}pers,~H.; Somaschini,~C.; Geelhaar,~L.
  \emph{Nanotechnology} \textbf{2016}, \emph{27}, 095601\relax
\mciteBstWouldAddEndPuncttrue
\mciteSetBstMidEndSepPunct{\mcitedefaultmidpunct}
{\mcitedefaultendpunct}{\mcitedefaultseppunct}\relax
\EndOfBibitem
\bibitem[Plissard \latin{et~al.}(2011)Plissard, Larrieu, Wallart, and
  Caroff]{Plissard2011}
Plissard,~S.; Larrieu,~G.; Wallart,~X.; Caroff,~P. \emph{Nanotechnology}
  \textbf{2011}, \emph{22}, 275602\relax
\mciteBstWouldAddEndPuncttrue
\mciteSetBstMidEndSepPunct{\mcitedefaultmidpunct}
{\mcitedefaultendpunct}{\mcitedefaultseppunct}\relax
\EndOfBibitem
\bibitem[Gibson \latin{et~al.}(2013)Gibson, Boulanger, and
  LaPierre]{Gibson2013b}
Gibson,~S.~J.; Boulanger,~J.~P.; LaPierre,~R.~R. \emph{Semiconductor Science
  and Technology} \textbf{2013}, \emph{28}, 105025\relax
\mciteBstWouldAddEndPuncttrue
\mciteSetBstMidEndSepPunct{\mcitedefaultmidpunct}
{\mcitedefaultendpunct}{\mcitedefaultseppunct}\relax
\EndOfBibitem
\bibitem[Rudolph \latin{et~al.}(2014)Rudolph, Schweickert, Mork{\"{o}}tter,
  Loitsch, Hertenberger, Becker, Bichler, Abstreiter, Finley, and
  Koblmüller]{Rudolph2014}
Rudolph,~D.; Schweickert,~L.; Mork{\"{o}}tter,~S.; Loitsch,~B.;
  Hertenberger,~S.; Becker,~J.; Bichler,~M.; Abstreiter,~G.; Finley,~J.~J.;
  Koblmüller,~G. \emph{Applied Physics Letters} \textbf{2014}, \emph{105},
  033111\relax
\mciteBstWouldAddEndPuncttrue
\mciteSetBstMidEndSepPunct{\mcitedefaultmidpunct}
{\mcitedefaultendpunct}{\mcitedefaultseppunct}\relax
\EndOfBibitem
\bibitem[Dalacu \latin{et~al.}(2009)Dalacu, Kam, {Guy Austing}, Wu, Lapointe,
  Aers, and Poole]{Dalacu2009}
Dalacu,~D.; Kam,~A.; {Guy Austing},~D.; Wu,~X.; Lapointe,~J.; Aers,~G.~C.;
  Poole,~P.~J. \emph{Nanotechnology} \textbf{2009}, \emph{20}, 395602\relax
\mciteBstWouldAddEndPuncttrue
\mciteSetBstMidEndSepPunct{\mcitedefaultmidpunct}
{\mcitedefaultendpunct}{\mcitedefaultseppunct}\relax
\EndOfBibitem
\bibitem[Kelrich \latin{et~al.}(2013)Kelrich, Calahorra, Greenberg, Gavrilov,
  Cohen, and Ritter]{Kelrich2013}
Kelrich,~A.; Calahorra,~Y.; Greenberg,~Y.; Gavrilov,~A.; Cohen,~S.; Ritter,~D.
  \emph{Nanotechnology} \textbf{2013}, \emph{24}, 475302\relax
\mciteBstWouldAddEndPuncttrue
\mciteSetBstMidEndSepPunct{\mcitedefaultmidpunct}
{\mcitedefaultendpunct}{\mcitedefaultseppunct}\relax
\EndOfBibitem
\bibitem[Gibson and LaPierre(2014)Gibson, and LaPierre]{Gibson2014}
Gibson,~S.~J.; LaPierre,~R.~R. \emph{Nanotechnology} \textbf{2014}, \emph{25},
  415304\relax
\mciteBstWouldAddEndPuncttrue
\mciteSetBstMidEndSepPunct{\mcitedefaultmidpunct}
{\mcitedefaultendpunct}{\mcitedefaultseppunct}\relax
\EndOfBibitem
\bibitem[Dubrovskii \latin{et~al.}(2015)Dubrovskii, Xu, Diaz, Plissard, Caroff,
  Glas, and Grandidier]{Dubrovskii2015b}
Dubrovskii,~V.~G.; Xu,~T.; Diaz,~A.; Plissard,~S.~R.; Caroff,~P.; Glas,~F.;
  Grandidier,~B. \emph{Nano Letters} \textbf{2015}, \emph{15}, 5580--5584\relax
\mciteBstWouldAddEndPuncttrue
\mciteSetBstMidEndSepPunct{\mcitedefaultmidpunct}
{\mcitedefaultendpunct}{\mcitedefaultseppunct}\relax
\EndOfBibitem
\bibitem[Ralston \latin{et~al.}(1986)Ralston, Wicks, and Eastman]{Ralston}
Ralston,~J.; Wicks,~G.~W.; Eastman,~L.~F. \emph{Journal of Vacuum Science \&
  Technology B: Microelectronics Processing and Phenomena} \textbf{1986},
  \emph{4}, 594--597\relax
\mciteBstWouldAddEndPuncttrue
\mciteSetBstMidEndSepPunct{\mcitedefaultmidpunct}
{\mcitedefaultendpunct}{\mcitedefaultseppunct}\relax
\EndOfBibitem
\bibitem[Fick(1855)]{Fick1851}
Fick,~A. \emph{Annalen der Physik und Chemie} \textbf{1855}, \emph{170},
  59--86\relax
\mciteBstWouldAddEndPuncttrue
\mciteSetBstMidEndSepPunct{\mcitedefaultmidpunct}
{\mcitedefaultendpunct}{\mcitedefaultseppunct}\relax
\EndOfBibitem
\bibitem[Ramdani \latin{et~al.}(2013)Ramdani, Harmand, Glas, Patriarche, and
  Travers]{Ramdani2013}
Ramdani,~M.~R.; Harmand,~J.-C.~C.; Glas,~F.; Patriarche,~G.; Travers,~L.
  \emph{Crystal Growth {\&} Design} \textbf{2013}, \emph{13}, 91--96\relax
\mciteBstWouldAddEndPuncttrue
\mciteSetBstMidEndSepPunct{\mcitedefaultmidpunct}
{\mcitedefaultendpunct}{\mcitedefaultseppunct}\relax
\EndOfBibitem
\bibitem[Casadei \latin{et~al.}(2013)Casadei, Krogstrup, Heiss, Rohr, Colombo,
  Ruelle, Upadhyay, Sorensen, Nygard, and {Fontcuberta i Morral}]{Casadei2012}
Casadei,~A.; Krogstrup,~P.; Heiss,~M.; Rohr,~J.~a.; Colombo,~C.; Ruelle,~T.;
  Upadhyay,~S.; Sorensen,~C.~B.; Nygard,~J.; {Fontcuberta i Morral},~A.
  \emph{Applied Physics Letters} \textbf{2013}, \emph{102}, 013117\relax
\mciteBstWouldAddEndPuncttrue
\mciteSetBstMidEndSepPunct{\mcitedefaultmidpunct}
{\mcitedefaultendpunct}{\mcitedefaultseppunct}\relax
\EndOfBibitem
\bibitem[K\"upers \latin{et~al.}(2017)K\"upers, Tahraoui, Lewis, Rauwerdink,
  Matalla, Kr\"uger, Bastiman, Riechert, and Geelhaar]{SAGtech}
K\"upers,~H.; Tahraoui,~A.; Lewis,~R.~B.; Rauwerdink,~S.; Matalla,~M.;
  Kr\"uger,~O.; Bastiman,~F.; Riechert,~H.; Geelhaar,~L. \emph{arXiv e-prints}
  \textbf{2017}, \emph{arXiv:1708.02454}\relax
\mciteBstWouldAddEndPuncttrue
\mciteSetBstMidEndSepPunct{\mcitedefaultmidpunct}
{\mcitedefaultendpunct}{\mcitedefaultseppunct}\relax
\EndOfBibitem
\end{mcitethebibliography}

\end{document}